\documentclass[%
twocolumn, superscriptaddress,
nofootinbib,
 amsmath,amssymb,
 aps,
pra,
]{revtex4}
\usepackage[all]{xy}
\usepackage{mathrsfs}
\usepackage{graphicx}
\usepackage{epstopdf}
\usepackage{dcolumn}
\usepackage{bm}
\usepackage{color} 
\usepackage{ctable}  
\usepackage{hyperref}
\usepackage{dsfont}
\usepackage{afterpage}

\newcommand{\bra}[1]{\langle #1 |}
\newcommand{\ket}[1]{\vert #1 \rangle}

\newcommand{\rhohat}{\hat{\rho}}

\newcommand{\nop}{\hat{n}}

\newcommand{\bop}{\hat{b}}
\newcommand{\bdop}{\hat{b}^{\dagger}}
\newcommand{\nhat}{\hat{n}}

\newcommand{\rC}{\tilde{\mathcal{C}}} 
\newcommand{\C}{\mathcal{C}} 
\newcommand{\im}{{\rm i}} 
\newcommand{\Sop}{{\hat{S}}} 
\newcommand{\Lso}{\mathcal{L}} 
\newcommand{\id}{{\hat{\textbf{1}}}} 
\newcommand{\Hop}{{\hat{H}}} 

\begin{document}
\title{Melting of the critical behavior of a Tomonaga-Luttinger liquid under dephasing}

        \author{Jean-S\'ebastien Bernier\footnote{These authors contributed equally to the work.}}
	\affiliation{Physikalisches Institut, University of Bonn, Nussallee 12, 53115 Bonn, Germany}
	
	\author{Ryan Tan\footnotemark[\value{footnote}]}
	\affiliation{Engineering Product Development Pillar, Singapore University of Technology and Design, 8 Somapah Road, 487372 Singapore}
		
	\author{Chu Guo} 
	\affiliation{Supremacy Future Technologies, Guangzhou 511340, China}           		
		
	\author{Corinna Kollath}
	\affiliation{Physikalisches Institut, University of Bonn, Nussallee 12, 53115 Bonn, Germany}
	
	\author{Dario Poletti}
	\email{dario\_poletti@sutd.edu.sg} 
	\affiliation{Science and Math Cluster, Singapore University of Technology and Design, 8 Somapah Road, 487372 Singapore}
	\affiliation{Engineering Product Development Pillar, Singapore University of Technology and Design, 8 Somapah Road, 487372 Singapore}

\begin{abstract}
Strongly correlated quantum systems often display universal behavior as, in certain regimes, their properties
are found to be independent of the microscopic details of the underlying system. An example of such a situation is
the Tomonaga-Luttinger liquid description of one-dimensional strongly correlated bosonic or fermionic systems.
Here we investigate how such a quantum liquid responds under dissipative dephasing dynamics and, in particular,
we identify how the universal Tomonaga-Luttinger liquid properties melt away. Our study, based on adiabatic elimination,
shows that dephasing first translates into the damping of the oscillations present in the density-density correlations, a behavior
accompanied by a change of the Tomonaga-Luttinger liquid exponent. This first regime is followed 
by a second one characterized by the diffusive propagation of featureless correlations as expected for an infinite temperature state. 
We support these analytical predictions by numerically exact simulations carried out using a number-conserving implementation
of the matrix product states algorithm adapted to open systems.
\end{abstract}

\maketitle
	
\section{Introduction}
In equilibrium many-body systems, the insensitivity of large scale macroscopic observables to the microscopic details of
the system is indicative of the emergence of universal behaviors~\cite{SachdevBook}. For instance, such universal behavior
is apparent in many one-dimensional systems where the low energy properties can be described by the Tomonaga-Luttinger liquid
theory both for fermions and bosons~\cite{GiamarchiBook}. Investigating the emergence of such behaviors is
not restricted to equilibrium correlated systems. In fact, understanding how universality arises in out-of-equilibrium
systems is both of fundamental and practical interest. For example, developing the next generation of quantum technologies will
very likely require a thorough understanding of the response of large many-body quantum systems to both unitary and dissipative
perturbations and drivings.    

In this context, one important question relates to the robustness of the universal behavior of a Tomonaga-Luttinger liquid
to environmental couplings. A direction pursued, for instance, in~\cite{BuchholdDiehl2015} 
where the evolution of a Tomonaga-Luttinger liquid in contact with a dephasing environment was considered using a Keldysh
functional approach~\cite{SiebererDiehl2016, KamenevBook}, and in~\cite{BacsiDora2019} where a Tomonaga-Luttinger liquid
coupled quadratically to a finite temperature bath was studied. These works underline that it is now possible
to consider the combined influence of universality and quantum dissipation on the non-equilibrium
dynamics of strongly correlated systems. Theoretical predictions made from newly available numerical and analytical methods can even
be tested in state-of-the-art ultracold atoms experiments. For example, ultracold atoms under the effect of dephasing were studied
in~\cite{PatilVengalattore2015, LuschenSchneider2017, BouganneGerbier2020}, and the dissipative dynamics of Rydberg
atoms were investigated in~\cite{MalossiMorsch2014, ValadoMorsch2016, GutierrezMorsch2017}.

On the theoretical side, one method has come to prominence 
to study many-body open quantum systems and relies
on adiabatically eliminating fast decaying or decohering modes (see for
instance~\cite{DurrRempe2009, Garcia-RipollCirac2009, PolettiKollath2012}). This so-called adiabatic elimination
technique was successfully applied to explain the nature of the
dephasing dynamics of a bosonic gas, and led to the prediction of interaction-induced
impeding. This phenomenon is characterized by the ``slowing down'' of a system relaxation dynamics
due to the interplay of interaction, kinetic and dissipation effects~\cite{PolettiKollath2012}. This 
impeding can take various forms as it can give rise to emerging power-law~\cite{PolettiKollath2012, CaiBarthel2013}
and stretched exponential~\cite{PolettiKollath2013} relaxations, and can even lead to the occurrence of
aging~\cite{SciollaKollath2015, WolffKollath2019}. In fact, the power-law relaxation dynamics predicted analytically by
adiabatic elimination in~\cite{PolettiKollath2012, PolettiKollath2013} has by now been observed experimentally
in~\cite{BouganneGerbier2020}. One should also note that impeded relaxation dynamics were also
predicted to occur in the presence of disorder~\cite{MedvedyevaZnidaric2016, EverestLevi2017} or kinetic
constraints~\cite{OlmosGarrahan2012, LesanovskyGarrahan2013}.

In addition, important progress has recently been made to optimize the matrix product
states (MPS) algorithm~\cite{White1992, Schollwoeck2011} to study open quantum systems.
While MPS has been used to investigate such dissipative systems for
several years~\cite{VerstraeteCirac2004, ZwolakVidal2004, Schollwoeck2011,Daley2014},
a number-conserving algorithm for open quantum systems was only implemented
recently (see e.g.~\cite{BonnesLaeuchli2014b, BernierKollath2018, WolffKollath2019}).
For a certain class of systems, this approach was found to significantly reduce the size of the matrices that need to be manipulated,
thus allowing one to consider larger systems.    

Here we use both adiabatic elimination and a number-conserving MPS algorithm for open systems to study the
relaxation dynamics of a Tomonaga-Luttinger liquid under the effect of dephasing.
From adiabatic elimination, we predict that the spatial decay of the correlations is altered from
its initial universal algebraic behavior and adopts the algebraic scaling expected for a non-interacting gas.
At later times, a second regime diffusively takes over, and is characterized by featureless correlations
corresponding to the infinite temperature state. We complement the adiabatic elimination study
with quasi-exact numerical simulations based on a number-conserving MPS algorithm. Using this method,
we study open quantum systems of up to $84$ sites and $21$ hard-core bosonic particles. Within the
time interval numerically reachable, we find our simulations to be consistent with the predictions
from adiabatic elimination. While the Tomonaga-Luttinger liquid long-range correlations appear to be
maintained over short times, the propagation of a dissipation-induced perturbation can be clearly seen.
We also detect a likely change of the algebraic exponent characterizing the spatial decay of the
Tomonaga-Luttinger correlations.

The manuscript is structured in the following manner: In Sec.~\ref{sec:Model}, we introduce the model
and describe the properties of the ground state of a system of hard-core bosons which we use as the initial condition for our study.
In Sec. \ref{sec:methods}, we provide an overview of the methods used: the adiabatic elimination technique
and the matrix product states algorithm with number conservation applied to the study of open systems.
In Sec.~\ref{sec:Results}, we present how the Tomonaga-Luttinger-like properties are affected by the dephasing
dynamics first using adiabatic elimination~\ref{sec:Results_AE}, and then, in Sec.~\ref{sec:Results_MPS},
we show how these results are supported by numerically exact MPS simulations.
We conclude in Sec.~\ref{sec:conclusions}.

\section{Model}\label{sec:Model}
We study the effects of dephasing on the short time dynamics of hard-core bosons in an optical lattice.
The Hamiltonian of the system is given by 
\begin{equation}\label{eq:bhmodel}
\hat{H} = -J \sum_{l=1}^{L-1} ( \hat{b}_l \hat{b}_{l+1}^\dag + \hat{b}_l^\dag \hat{b}_{l+1})  + V \sum_{l=1}^{L-1} \hat{n}_l \hat{n}_{l+1},
\end{equation}
where $J$ is the hopping magnitude, $V$ is the strength of the nearest-neighbour repulsion, $L$ is the number of sites.
Here we wrote down the Hamiltonian for open boundary conditions as will be considered in our numerical treatment using MPS
methods. $\hat{b}_l^{}$ and $\hat{b}_l^\dagger$ respectively annihilate and create a hard-core boson at site $l$.
The operator $\hat{n}_l=\hat{b}^\dagger_l\hat{b}_l$ counts the number of hard-core bosons at site $l$, and due to the hard-core bosons
conditions $\hat{b}^\dagger_l\hat{b}^\dagger_l=0$, $\hat{n}_l=0,\;1$. We consider the evolution of the density operator $\rhohat$ to
be described by a Lindblad master equation~\cite{BreuerPetruccione2002, GoriniSudarshan1976, Lindblad1976} with local dephasing on all sites of the system, 
\begin{equation}
\frac{d{\rhohat}}{dt} = -\frac{\im}{\hbar} [\hat{H},\rhohat] + \mathcal{D}(\rhohat), \label{eq:vonNeumann} 
\end{equation}
where the dissipator $\mathcal{D}(\rhohat)$ takes the form
\begin{equation}
\mathcal{D}(\hat{\rho}) = \gamma \sum_{l=1}^L 2 \nhat_l \hat{\rho} \hat{n}_l - \hat{n}_l^2 \rhohat - \rhohat \hat{n}_l^2. \label{eq:dissipator}  
\end{equation}
The jump operators $\hat{n}_l$, being Hermitian, drive the system towards the infinite temperature state which, given the interplay
between the Hamiltonian and the dissipator, is the only steady state of the system. In fact, this dissipator is relevant to the 
study of ultracold atoms under the effect of fluorescent scattering~\cite{PichlerZoller2010, GerbierCastin2010},
and was also realized experimentally~\cite{BouganneGerbier2020, PatilVengalattore2015, LuschenSchneider2017}.

We prepare initially the system in the ground state of the Hamiltonian, Eq.~(\ref{eq:bhmodel}), and then switch on the dissipation
at $t > 0$. For the chosen fillings, $\bar{n}=\sum_l \langle \hat{n}_l\rangle/L$, and interaction strength, $0 \le V \le J$,
the initial state is typically well described as a Tomonaga-Luttinger liquid~\cite{GiamarchiBook, HikiharaFurusaki2001}.

In order to facilitate the use of the many-body adiabatic elimination method, we rewrite the lattice system of hard-core bosons
confined to an optical lattice and subjected to local dephasing noise
to the spin-$1/2$ XXZ chain coupled to dephasing. Using the mapping,
\begin{align}
  \bdop_l &\rightarrow \Sop^{+}_l, \nonumber \\
  \bop_l &\rightarrow \Sop^{-}_l, \nonumber \\
  \nop_l - \frac{1}{2} &\rightarrow \Sop^{z}_l, \nonumber
\end{align}
where $\Sop^\alpha_l$ is the $\alpha$-direction spin operator at site $l$, together with the transformation
\begin{align}
  \Sop^{+}_l &\rightarrow (-1)^l \Sop^{+}_l, \nonumber \\
  \Sop^{-}_l &\rightarrow (-1)^l \Sop^{-}_l, \nonumber \\
  \Sop^z_l &\rightarrow \Sop^z_l, \nonumber
\end{align}
the evolution of the density operator $\hat{\rho}$ can now be described by the Lindblad master equation
\begin{equation}
\frac{d{\rhohat}}{dt} = -\frac{\im}{\hbar} [\hat{H}_\text{XXZ},\rhohat] + \mathcal{D}_\text{Z}(\rhohat)~. \nonumber
\end{equation}
Here, the XXZ Hamiltonian with periodic boundary conditions is
\begin{align}
\hat{H}_{\text{XXZ}} = \sum_{l=1}^{L} \left[J \left( \Sop^{+}_l \Sop^{-}_{l+1} + \Sop^{+}_{l+1} \Sop^{-}_{l} \right) + V \Sop^z_l \Sop^z_{l+1} \right],\nonumber
\end{align}
where $J$ and $V$ are the exchange couplings along the different spin directions, $L$ is the length of the chain, and
$\Sop^\alpha_{L+1} = \Sop^\alpha_{1}$. The dissipator takes the form 
\begin{align}
\mathcal{D}_\text{Z}(\rhohat) &= 2 \gamma \sum_{l=1}^L \left(\Sop^z_l \rhohat \Sop^z_l  - \frac{1}{4} \rhohat \right),\nonumber
\end{align}
where $\gamma$ is the dissipation strength. In this notation the dissipation induces
spin fluctuations.

As we consider a situation where the system is initially prepared as a Tomonaga-Luttinger liquid, we briefly
discuss the structure of the correlations associated with this state as they are used as
inputs within adiabatic elimination. In this liquid, for $L \rightarrow \infty$, the spin correlations along the $z$-direction 
decay algebraically as
\begin{align}
\langle  \Sop^z_{l}(0) \Sop^z_{l+d}(0) \rangle &= \left(\bar{n} - \frac{1}{2} \right)^2 + A_z (-1)^{d} \frac{\cos(q d)}{|d|^{1/\eta}} \nonumber \\
                                 &~~~~~~- \frac{1}{4 \pi^2 \eta |d|^2} \label{eq:TLcorrel}
\end{align}
where $q = 2\pi \left(\bar{n} - \frac{1}{2}\right)$, $\eta = 1/(2K)$,
with $K$ the Tomonaga-Luttinger parameter, and $A_z$ is a function of
the interaction strength $V$~\cite{HikiharaFurusaki2001}. As for finite interaction strengths, $V > 0$, the Tomonaga-Luttinger
parameter $K < 1$, the term in $d^{-1/\eta} = d^{-2K}$ dominates. The correlations present oscillations whose period depends
on the filling, and the leading corrections go as $d^{-2}$. These spin correlations can be mapped back to the density-density
correlations as
\begin{align}
  \langle \Sop^z_l \Sop^z_{l+d} \rangle \rightarrow \langle \nop_l \nop_{l+d} \rangle - \bar{n} + \frac{1}{4}. \nonumber
\end{align}

Theoretically, this dissipative system has been the subject of a number of studies. The evolution of various
equal-time correlations were investigated in~\cite{CaiBarthel2013} whereas two-time density-density correlations,
or equivalently spin-spin correlations, were considered in~\cite{WolffKollath2019}. 

\section{Methods} \label{sec:methods}  
In this work we employ two complementary approaches. First, we use an approximate approach based on many-body
adiabatic elimination~\cite{DurrRempe2009, Garcia-RipollCirac2009, PolettiKollath2012} to build up
an analytical understanding of the correlation propagation regimes encountered at intermediate to long times. This method
is described in Sec.\ref{ssec:adel}. While adiabatic elimination provides deep insights into the evolution
of this correlated system under the effect of dephasing, it does not correctly capture the initial propagation dynamics and
does not assign the appropriate timescales to the different regimes. To fill some of these gaps, we
compute the short to intermediate time evolution of our system using a matrix product state (MPS) algorithm~\cite{Schollwoeck2011}
for open systems. This quasi-exact numerical method
allows for accurate estimations of the computational error which, in principle, can be made arbitrarily
small. We describe this approach in Sec.\ref{ssec:numerics}. 

\subsection{Adiabatic Elimination} \label{ssec:adel}    
To gain analytical insights into the evolution of the system, we employ many-body adiabatic elimination.
This method relies on the observation that, for sufficiently large times and irrespective of the system parameters,
the dissipation-free subspace will be reached. While this subspace is highly degenerate with respect to the dissipator, the
Hamiltonian lifts this degeneracy. Performing adiabatic elimination helps identifying how
virtual excitations around the dissipation-free subspace affect the system dynamics. 

For the system considered here, the dissipation-free subspace formed by all $\rhohat_0$ for which $\mathcal{D}_\text{Z}(\rhohat_0) = 0$,
takes the form $\rhohat_0 = \sum_{\vec{\sigma}} \rho_{0,\vec{\sigma}} \vert \vec{\sigma}\rangle \langle \vec{\sigma}\vert$
where the different spin configurations are labeled within the $z$-component basis such
that $\vec{\sigma} = (\sigma_1, \sigma_2, \cdots, \sigma_L)$ with $\sigma_l = \pm 1/2$. Adiabatic elimination
allows one to describe the effective dynamics within this subspace by considering the effect induced by
virtual excitations. As described in~\cite{WolffKollath2019}, for times larger than $1/\gamma$, the evolution of
the different elements of the density matrix is effectively described by the set of differential equations
\begin{align}
  &\frac{\partial \rho_{0,\vec{\sigma}}}{\partial t} =
  \sum_{l=1}^{L}\frac{J^2 \gamma}{\left[\left(V \alpha_{\vec{\sigma},l} \right)^2 + \left(\hbar\gamma\right)^2\right]}~\delta_{\sigma_l, \bar{\sigma}_{l+1}}
  \left( \rho_{0, \vec{\sigma}_l} - \rho_{0, \vec{\sigma}}\right), \nonumber
\end{align}
where $\alpha_{\vec{\sigma},l} = (\sigma_{l-1} \sigma_{l} +  \sigma_{l+1}  \sigma_{l+2})$, $\vec{\sigma}_l$ is
the spin configuration $\vec{\sigma}$ with swapped spins at site $l$ and $l+1$.

We then use this set of differential equations for $\rho_0$ to write down a set of coupled differential
equations for the spin correlations $\C_{l,l+d}(t) = \langle \Sop^z_{l}(t) \Sop^z_{l+d}(t) \rangle$. Together with periodic boundary
conditions, these equations, valid for $\hbar\gamma \gg V$ (as these expressions were obtained by formally taking the limit
$V \rightarrow 0$), take the form
\begin{align}
  \frac{\partial}{\partial t}\C_{j,j \pm 1} &= D \left(\C_{j \mp 1,j \pm 1} + \C_{j,j \pm 2} - 2~\C_{j,j \pm 1} \right), \nonumber \\
  \frac{\partial}{\partial t}\C_{j,j+d} &= D \left(\C_{j+1,j+d} + \C_{j-1,j+d} + \C_{j,j+d+1} \right. \nonumber \\
  & \qquad \left. +~\C_{j,j+d-1} - 4~\C_{j,j+d} \right),~~~~\text{for $|d|>1$}, \nonumber
\end{align}
where $D = J^2/(\hbar^2\gamma)$ and we dropped here the time from $\C_{l, l+d}(t)$ to simplify the notation.
As the system is initially prepared as a Tomonaga-Luttinger liquid, the correlations are translationally invariant,
$\C_{d}(t) = \C_{j,j+d}(t)$, such that the system of equations can be simplified to
\begin{align}
  \frac{\partial}{\partial t}\C_{\pm 1} &= 2 D \left(\C_{\pm 2} - \C_{\pm 1} \right), \label{eq:diffeqtransinv} \\
  \frac{\partial}{\partial t}\C_{d} &= 2 D \left(\C_{d+1} + \C_{d-1} - 2~\C_{d} \right), \nonumber
\end{align}
where, for the second equation, $-L/2+1~\leq~d~<~-1$ or $1~<~d~\leq~L/2$.  

To solve this system of differential equations for which $\C_{d}(t) = \C_{-d}(t)$, it is convenient
to redefine the correlations such that the evolution for all distances is described by
differential equations of the same form. To do so, we redefine the correlations as
$\rC_d(t) = \C_d(t)$ for $1 \leq d \leq L/2$ and $\rC_{d+1}(t) = \C_d(t)$ for $-L/2+1 \leq d \leq -1$
implying that $\rC_d(t) = \rC_{-d+1}(t)$ for $1 \leq d \leq \frac{L}{2}$. One
can then write down a diffusion equation for $\rC_{d}$ with diffusion constant $D$
\begin{align}
  \frac{\partial}{\partial t}\rC_{d} =  2D \left(\rC_{d+1} + \rC_{d-1} - 2~\rC_{d} \right) \nonumber
\end{align}
valid for $-L/2+2\leq d \leq L/2$.

This differential equation can be solved analytically in terms of the modified Bessel functions $I_n(x)$, and has for solution
\begin{align}
  \rC_{d}(t) &= \left(\bar{n} - \frac{1}{2} \right)^2 + e^{-4 D t} \left(\sum_{d'=1}^{L/2}~\mathcal{C}_{d'}(0)~I_{d'-d}(4 D t) \right. \nonumber \\
  &~~~~~~~~~~~ \left. + \sum_{d'=1}^{L/2-1}~\mathcal{C}_{d'}(0)~I_{d'+d-1}(4 D t) \right), \label{eq:fullanalytical}
\end{align}
where $\mathcal{C}_d(0) = \langle  S^z_{l}(0) S^z_{l+d}(0) \rangle$ are the spin correlations in the $z$-direction. 
For $d \ge 1$, in the limit where $L \gg 1$, the correlations can be written as
\begin{align}
  & \C_{d}(t) = \left(\bar{n} - \frac{1}{2} \right)^2 \label{eq:fullanalyticalnontilde} \\
  &~~~ + e^{-4 D t} \left(\sum_{d'=1}^{\infty}~\mathcal{C}_{d'}(0)~\left[I_{d'-d}(4 D t) + I_{d'+d-1}(4 D t)\right] \right). \nonumber
\end{align}
Using the initial conditions given by Eq.~(\ref{eq:TLcorrel}), it is instructive to consider two limits.
For $d > \sqrt{4 D t}$, we find that these correlations are well approximated
by the expression
\begin{align}
  \C_{d}(t) &= \left(\bar{n} - \frac{1}{2} \right)^2 + A_z \frac{(-1)^d}{d^{1/\eta}}~e^{-4 D t (1 + \cos q)} \cos(q d) \nonumber \\
  &~~~~~ - \frac{1}{4\pi^2\eta} \frac{1}{d^2},  \label{eq:cdapprox1}
\end{align}
where $A_z$ is a constant. 
This result suggests that the algebraic decay characterized by the Tomonaga-Luttinger liquid exponent and the associated
oscillations are damped out on a timescale set by $4 D (1 + \cos q)$. However, one should note that the validity of these
results at short times need to be confirmed using more robust methods as, initially, the system might not be close
enough to the dissipation-free subspace. For longer times, as only the term in $d^{-1/\eta}$ is damped, the correlations will remain algebraic for
distances larger than $\sqrt{4 D t}$, but their scaling exponent will change. In fact, irrespective of the initial interaction
strength $V$, there exists a certain intermediate regime, in distance and time, where the correlations will scale as $d^{-2}$.

At very long times, the correlations become uniform over all distances and their value varies as 
\begin{align}
  \C_d(t) &= \left(\bar{n} - \frac{1}{2} \right)^2 + \frac{1}{\sqrt{2 \pi D t}}~S_0, \label{eq:cdapprox2}
\end{align}
where
\begin{align}
 S_0 = \frac{A_z}{2} \left(\text{Li}_{\frac{1}{\eta}}(-e^{-iq}) + \text{Li}_{\frac{1}{\eta}}(-e^{iq}) \right) - \frac{1}{24 \eta}, \nonumber
\end{align}
with $\text{Li}_n(x)$, the polylogarithm function. One sees here that the correlations along the $z$-direction approach
their final value $(\bar{n} - 1/2)^2$ following a scaling $t^{-1/2}$. We detail further the evolution of these
correlations in Sec.~\ref{sec:Results_AE} where we discuss the results obtained within the adiabatic elimination formalism.

\subsection{Matrix Product States} \label{ssec:numerics}    
Simulating exactly the time evolution of interacting lattice bosons under the effects of dephasing would require a very large amount of memory since we are dealing with density matrices which require the square of the elements needed when considering wavefunctions. 
In order to circumvent this problem, we use a number-conserving MPS algorithm adapted to dissipative systems. 
We reshape the density matrix to a vector which we then represent within the MPS formalism (for the first works in this direction, although without number conservation, see \cite{VerstraeteCirac2004, ZwolakVidal2004}). 
Methods based on matrix product states rely on rewriting a quantum states as \cite{Schollwoeck2011}
\begin{align}\label{eq:psimps}   
\ket{\psi} = \sum_{\sigma_1 \dots \sigma_L} F^{\sigma_1} \dots F^{\sigma_L} \ket{\sigma_1 \dots \sigma_L}, 
\end{align}
where $\sigma_l$ represents the physical degree of freedom on the $l$-th site. 
For a system with a $U(1)$ symmetry, e.g. particle number conservation, each $F^{\sigma_l}$ can be taken as   
a matrix $F^{\sigma_l}_{(a_l,\alpha_l),(a_{l+1},\alpha_{l+1})}$, with auxiliary indices $a_l$ and $a_{l+1}$, and labelled by quantum numbers $\alpha_l$ and $\alpha_{l+1}$. In the case of a sufficiently large matrix dimension, the representation of the state is exact. The idea which makes the MPS methods feasible is the truncation of the maximum sizes of the auxiliary dimensions using singular value decompositions, while the quantum numbers are constrained by $\alpha_l + \sigma_l = \alpha_{l+1}$, with $\alpha_L = N$, where $N$ is the total number of atoms.

In order to deal with density matrices, we rewrite $\rhohat=\ket{\psi}\bra{\psi}$ as a vector $\ket{\rho}\rangle$, and then in close analogy as for the vector that describes a quantum state, we can write  
\begin{align} \label{eq:denmatmps}
\ket{\rho}\rangle = \sum_{\substack{\sigma_1 \dots \sigma_L \\ \sigma_1' \dots \sigma_L'}} M^{\sigma_1\sigma_1'} \dots M^{\sigma_L \sigma_L'} \ket{\sigma_1\sigma_1' \dots \sigma_L\sigma_L'}\rangle,  
\end{align}
where now the system has a $U(1)\otimes U(1)$ symmetry. The $l-$th site is, more precisely, represented by the tensor $M^{\sigma_l\sigma_l'}_{(b_l,\alpha_l,\alpha_l'),(b_{l+1},\alpha_{l+1},\alpha_{l+1}')}$ where $\alpha_l + \sigma_l = \alpha_{l+1}$ and $\alpha_l' + \sigma_l' = \alpha_{l+1}'$ to ensure number conservation, while $b_l$ and $b_{l+1}$ are auxiliary indices.
Here $\sigma_l'$ indicates the physical degree of freedom on the $l$-th site of the bra of the density operator. 
The average value of a generic observable $\hat{O}$ is computed as $\langle \hat{O} \rangle={\rm tr}(\hat{O} \hat{\rho})$, which can be rewritten for the vectorized $|\rho\rangle\rangle$ as $\langle \hat{O} \rangle = \langle\langle \mathds{1}| \hat{O} |\rho \rangle\rangle $, where $|\mathds{1} \rangle\rangle$ is the vectorized identity operator labelled with the same symmetries as $|\rho\rangle\rangle$.   

At the initial time, the MPS $M^{\sigma_l\sigma_l'}$ for the density matrix are converted from the MPS in Eq.(\ref{eq:psimps}) which represent the ground state by the  following relation
\begin{align}
M^{\sigma_l \sigma'_l}_{(b_l,\alpha_l,\alpha_l'),(b_{l+1},\alpha_{l+1},\alpha_{l+1}')} =& F^{\sigma_l}_{(a_l,\alpha_l),(a_{l+1},\alpha_{l+1})} \nonumber\\
&\otimes F^{\sigma'_l}_{(a'_l,\alpha'_l),(a'_{l+1},\alpha'_{l+1})}. \nonumber 
\end{align}
In our calculation we consider a bond dimension $\chi$, given by the sum of all the local auxiliary indices $b_l$, of up to $8000$ levels, and we remove singular values smaller than $\varepsilon=10^{-7}$.  
In order to ensure that the ground state is well represented after conversion to the density operator $\rhohat$, we have evolved $\rhohat$ in absence of dephasing, i.e. $\gamma = 0$, to a time $Jt/\hbar = 1$ and verified that observables such as local density and local fluctuations remain constant up to an absolute accuracy of $10^{-9}$. 

This way of setting up the MPS representation of $\ket{\rho}\rangle$ allows one to write Eqs.(\ref{eq:vonNeumann}) and (\ref{eq:dissipator})
in order to calculate the time-evolution using 
\begin{align}
\frac{d\ket{\rho}\rangle}{dt} = \Lso \ket{\rho}\rangle \label{eq:vonNeumannVec} 
\end{align}
with the superoperator $\Lso$ given by  
\begin{align} 
\Lso &= -\frac{\im}{\hbar} \left( \Hop \otimes \id - \id \otimes \Hop \right) \nonumber \\ 
& + \gamma \sum_l \left( 2 \nop_l\otimes \nop_l - \nop_l^2 \otimes \id - \id \otimes {\nop_l^2} \right). \label{eq:Lso}  
\end{align} 
Note that the operators $\Hop$ and $\nop_l$ are symmetric, and hence we do not need to take the
transpose of the operators which would be acting on the bra. 

Given the representation chosen for $\ket{\rho\rangle}$ in Eq.(\ref{eq:denmatmps}), with MPS taking into account the
local indices of the ket and bra, respectively $\sigma_l$ and $\sigma_l'$, Eqs.(\ref{eq:vonNeumannVec}) and (\ref{eq:Lso})
result in operators acting only locally, or only coupling two nearest neighboring sites.  
We can thus decompose $\Lso = {\Lso}_E + {\Lso}_O$ where $\Lso_E$ and $\Lso_O$ act, respectively, only on the even or the odd bonds.
We can then approximate the evolution of the density matrix over a small time interval $dt$ using a fourth-order
Trotter decomposition~\cite{JankeSauer1992}, which can be implemented using standard time-evolution MPS techniques,
see for instance~\cite{Schollwoeck2011}.  

\section{Results}\label{sec:Results}

\subsection{Dephasing dynamics within adiabatic elimination} \label{sec:Results_AE}
Investigating within adiabatic elimination the dissipative evolution of the
density-density correlations, or correspondingly, for the XXZ Hamiltonian, the spin
correlations along the $z$-direction, we identify two distinct regimes. Considering
Eq.~(\ref{eq:fullanalyticalnontilde}), one sees that these correlations are functions of
the dimensionless time $D t$ where $D$ is the diffusion constant defined earlier
as $D = J^2/(\hbar^2 \gamma)$. We therefore use this quantity to delineate the different
regimes. While adiabatic elimination is unlikely to capture the exact timescales as
it notoriously underestimates the coherence initially present in the system, we can be fairly confident
that for $t \gg 1/\gamma$, or in other words $D t \gg \left[J/(\hbar \gamma)\right]^2$, this approach
provides a faithful description of the dynamical regimes occurring under the effect of dephasing.

\begin{figure}[h!]
\includegraphics[width=0.9\linewidth]{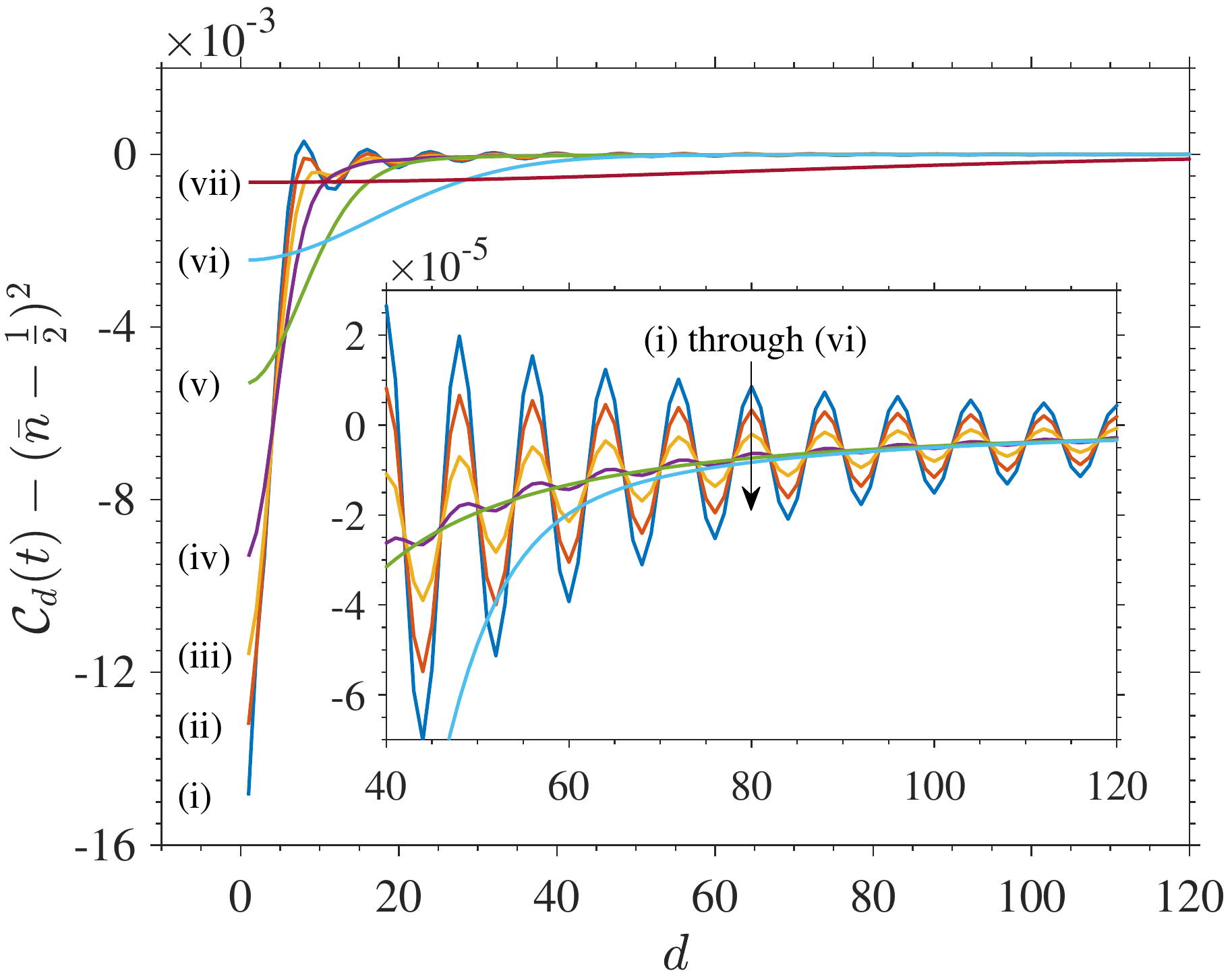}
\caption{Spin correlations along the $z$-direction, $\mathcal{C}_d(t)$, for a system of length $L = 2000$, filling $\bar{n} = 1/8$,
  initial interaction strength $V = J$, and dissipative strength $\hbar \gamma = 5 J$ corresponding to $\hbar^2 D = 0.2 J^2$.
  The curves represent  (i) $D t = 0$, (ii) $D t = 0.34$,
  (iii) $D t = 0.95$, (iv) $D t = 2.44$, (v) $D t = 11.24$, (vi) $D t = 62.98$, (vii) $D t = 995.54$. Inset: same labelling
  as in the main figure, curves (i) to (vi) are ordered from top to bottom, curve (vii) is not shown.
  The spatial oscillations of the Tomonaga-Luttinger correlations are damped under the effect of dephasing.
  The correlations are obtained by numerically solving the set of
  differential equations, Eqs.~(\ref{eq:diffeqtransinv}).}\label{fig:earlytimeslinear}
\end{figure}

For short to intermediate dimensionless times, we find that the algebraic decay of the
correlations is maintained for all distances, albeit with decreased amplitude and
changing exponent. Considering Eq.~(\ref{eq:cdapprox1}),
valid for $d > \sqrt{4\pi D t}$, one can see that for small $D t$ the term in $d^{-1/\eta}$ is
only weakly damped such that the correlations are still dominated by the same algebraic term, and
continue to present oscillations albeit with reduced amplitude. As time progresses, the window
over which Eq.~(\ref{eq:cdapprox1}) does not apply widens and, in fact, the initial
correlation structure completely melts away in this region. For $d \le \sqrt{4\pi D t}$,
the correlations adopt a flat plateau-like structure described by Eq.~(\ref{eq:cdapprox2})
and the plateau height decreases in time as $t^{-1/2}$. In contrast, for $d > \sqrt{4\pi D t}$,
the correlations still decay algebraically but with a different exponent: the first term is completely
damped out due to dephasing while the second term, in $d^{-2}$, remains unaffected. We present
in more details these two regimes below using different graphical representations.

Focusing first on the short to intermediate distance behavior, one sees from Fig.~\ref{fig:earlytimeslinear} 
that the spatial oscillations present in the initial Tomonaga-Luttinger liquid correlations
are damped under the effect of dephasing. For larger $D t$, at small distances one can notice a slow
build-up as the correlations approach their final value $(\bar{n}^2 - 1/2)^2$. In Fig.~\ref{fig:earlytimeslinear}, this
build-up and the melting of the algebraically decaying character is seen to propagate to larger correlation distances
with increasing time.

\begin{figure}[h!]
\includegraphics[width=0.9\linewidth]{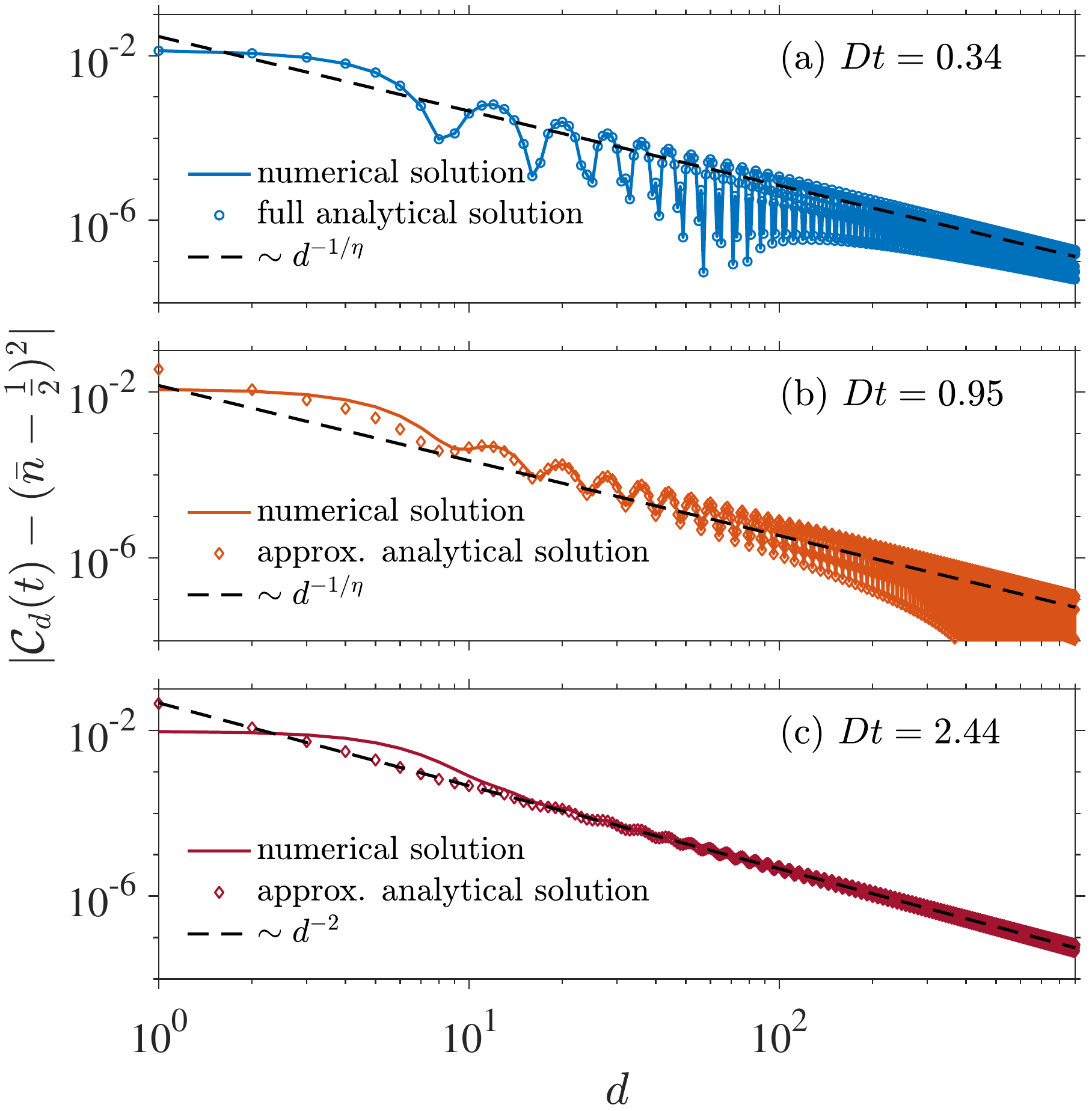}
\caption{Spatial scaling of the spin correlations along the $z$-direction, $\mathcal{C}_d(t)$, for short to
  intermediate times. The same system parameters as in Fig.~\ref{fig:earlytimeslinear} are used. In (a), the numerical solution
  (solid line) is compared to the analytical solution (circle markers), Eq.~(\ref{eq:fullanalytical}). In (b) and (c),
  the numerical solution (solid line) is compared to the simplified expression, Eq.~(\ref{eq:cdapprox1})
  (diamond markers).}\label{fig:earlytimeslog}
\end{figure}

Useful information about the evolution of the system can be extracted by considering the correlations
over longer distances. At early times, as shown in Fig.~\ref{fig:earlytimeslog} (a) and (b), the correlations
in the tail scale as $d^{-1/\eta}$ as expected from Tomonaga-Luttinger theory. However, for increasing times,
as seen in (c), the scaling changes to $d^{-2}$. This behavior is in
agreement with Eq.~(\ref{eq:cdapprox1}), an approximated expression for $\mathcal{C}_d(t)$, valid for $d > \sqrt{4 \pi D t}$ which,
as explained earlier, highlights that only the $d^{-1/\eta}$ term is exponentially damped while the $d^{-2}$
is unaffected. In Fig.~\ref{fig:earlytimeslog}, we further compare the correlations obtained from numerically solving
the set of differential equations, Eqs.~(\ref{eq:diffeqtransinv}), to two analytical solutions Eqs.~(\ref{eq:fullanalytical}) and
(\ref{eq:cdapprox1}). In panel (a), one sees
that the correlations obtained from the full analytical solution, Eq.~(\ref{eq:fullanalytical}), (circle markers)
agree perfectly with the numerical solution (solid line). Whereas, in panels (b) and (c), one can
see that the simplified analytical expression, Eq.~(\ref{eq:cdapprox1}), (diamond markers) captures very well the correlations
for $d > \sqrt{4 \pi D t}$.

\begin{figure}[h!]
\includegraphics[width=0.9\linewidth]{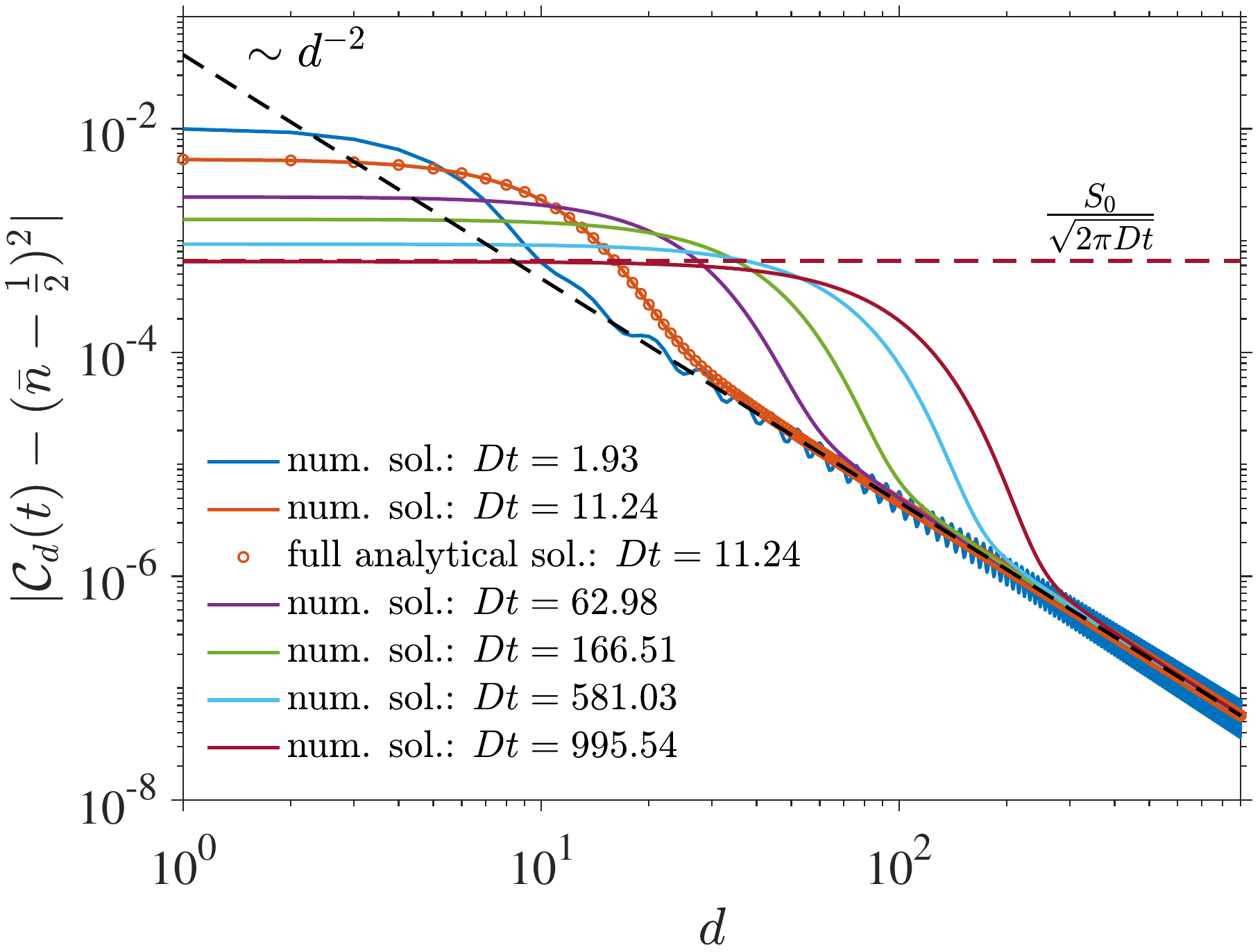}
\caption{Appearance, at longer times, of a second regime characterized by spatially uniform correlations.
  The same system parameters as in Fig.~\ref{fig:earlytimeslinear} are used. The curves are labelled in order of appearance at $d=1$ from top
  to bottom. The value of the correlation plateau scales as $t^{-1/2}$, and beyond the propagation front
  the correlations scale as $d^{-2}$. For $Dt = 11.24$, one sees that, in this second regime, the correlations obtained
  from Eq.~(\ref{eq:fullanalytical}) (circle markers) also agree perfectly with the numerical
  solution (solid line).}\label{fig:longtimeslog}
\end{figure}

For larger $D t$,  a second regime sets in. In this regime, the correlations initial algebraic decay
is completely obliterated for $d \le \sqrt{4 \pi D t}$. For these distances, the amplitude of the correlations
becomes spatially uniform and is only a function of time. As shown in Fig.~\ref{fig:longtimeslog}, the range of distances
over which the correlations are featureless increases as a function of time, and their amplitude decreases as $t^{-1/2}$ towards
their steady-state value of $(\bar{n} - 1/2)^2$ as described by Eq.~(\ref{eq:cdapprox2}). Beyond this propagation
front, the spin correlations along the $z$-direction still scale as $d^{-2}$ as found in the previous regime.

\begin{figure}[h!]
\includegraphics[width=0.9\linewidth]{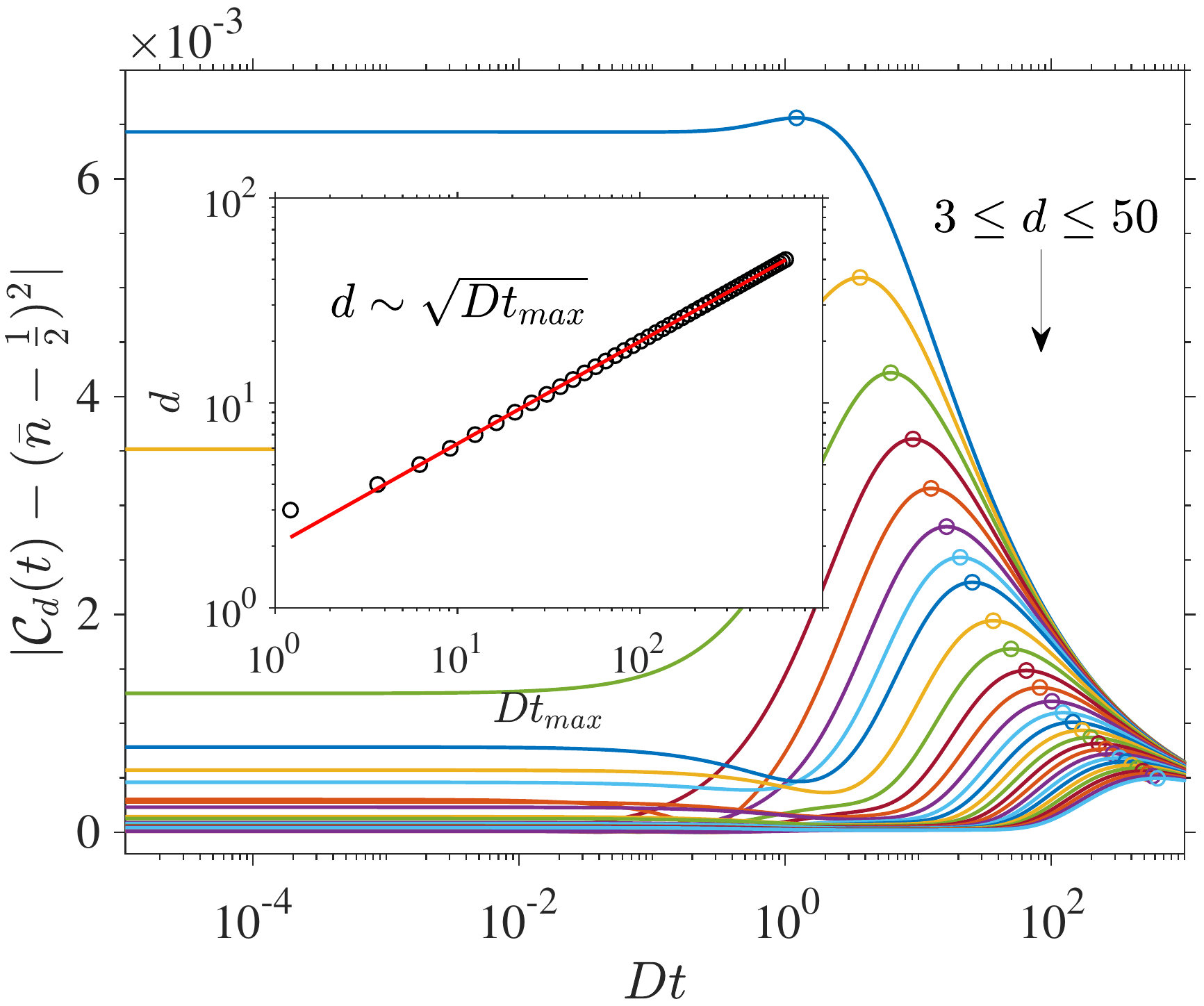}
\caption{Evolution of the spin correlations along the $z$-direction, $\mathcal{C}_d(t)$, as a function
  of time for $3 \le d \le 50$ (from the right, top to bottom). The same system parameters as
  in Fig.~\ref{fig:earlytimeslinear} are used. For each distance, a circle marks the time, $t_\text{max}$, at which the correlations
  are maximum. This maximum corresponds to the passage of the propagation front separating the regions displaying
  uniform and algebraically decaying correlations. Inset: the front propagates diffusively
  as $d \sim \sqrt{D t_\text{max}}$.}\label{fig:diffusive}
\end{figure}

As alluded to earlier, when considering Fig.~\ref{fig:earlytimeslinear}, the algebraically decaying character of the
correlations melts away over larger distances as time progresses. We want to understand here the mechanism by which the initial algebraically
decaying correlation structure is being replaced by spatially uniform correlations under the effect of dissipation.
To do so, one can notice from Fig.~\ref{fig:longtimeslog} that, for a given distance, $|\mathcal{C}_d(t) - (\bar{n} - 1/2)^2|$ is maximal
around the time when the propagation front, the flat region boundary, reaches this distance.
The time-dependence of the spin correlations along the $z$-direction
for $3 \le d \le 50$ is shown in Fig.~\ref{fig:diffusive}. The correlations peak at the passage of the front, at a value we denote
as $t_\text{max}$, and then decay algebraically, as $t^{-1/2}$, towards their steady-state value, $(\bar{n} - 1/2)^2$. Plotting $d$ as
a function of $D t_\text{max}$, as presented in the inset of Fig.~\ref{fig:diffusive}, one sees that the front propagates as
$d \sim \sqrt{D t_\text{max}}$. This result indicates that the second regime, characterized by featureless correlations, propagates
diffusively through the system. In other words, the regime characterized by algebraically decaying correlations disappears
through the action of a diffusive process.

\subsection{Dephasing dynamics within matrix product states}\label{sec:Results_MPS}

As explained in the previous section, adiabatic elimination underestimates the coherence initially present in the
correlated system. Consequently, in order to put our previous results on a more solid ground, particularly the regime identified at short
times, we conduct full-fledged MPS simulations on systems with fillings $\bar{n} = 1/8$, $11$ bosons on $88$ sites,
and $\bar{n} = 1/4$, $21$ bosons on $84$ sites, and focus on the evolution of the density profile and density-density
correlations.

\begin{figure}[h!]
  \includegraphics[width=0.9\linewidth]{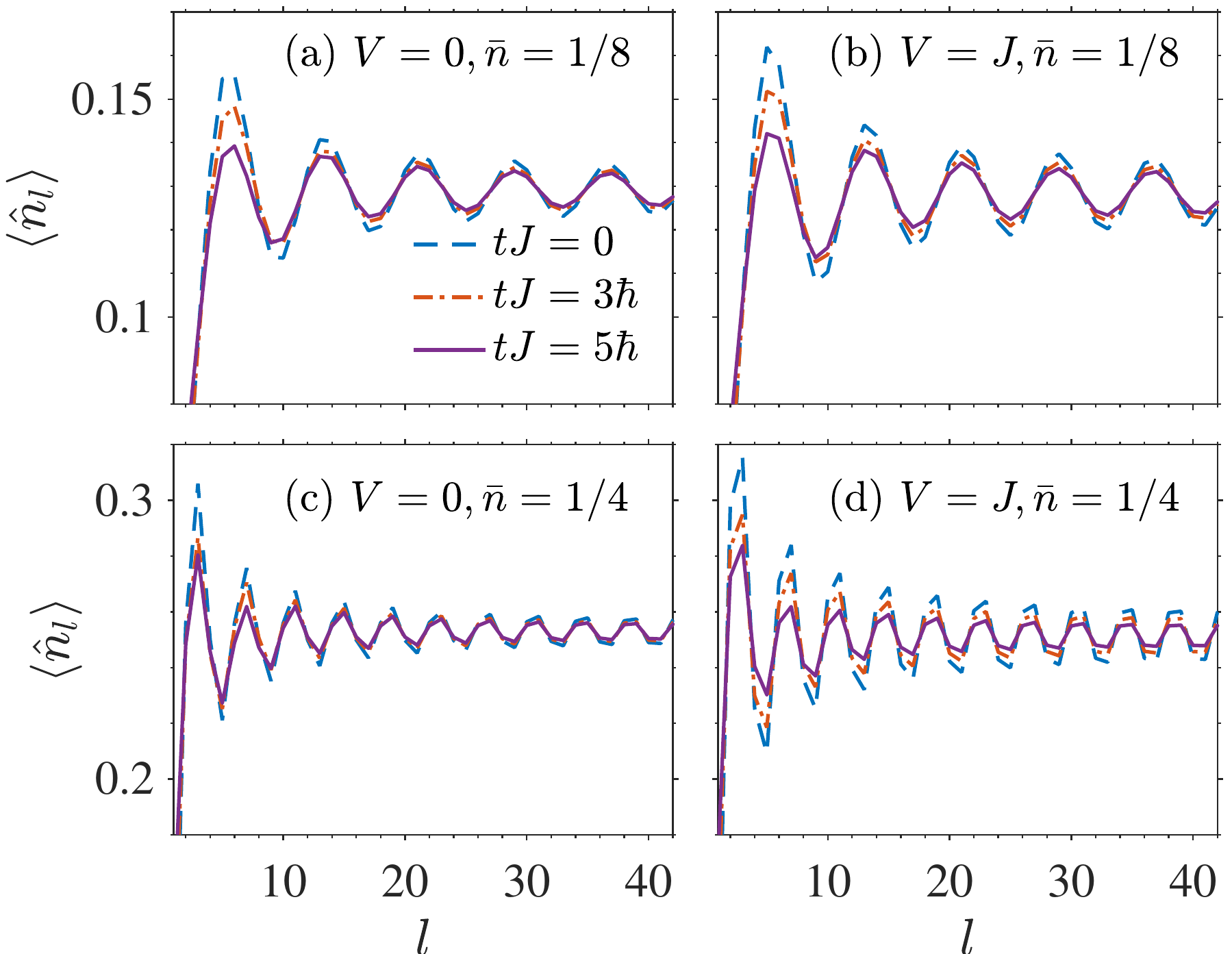}
  \caption{Local density $\langle \hat{n}_l \rangle$ versus site number $l$, for different times: $tJ = 0$ (blue dashed line),
    $tJ = 3 \hbar$ (orange dash-dotted line) and $tJ = 5 \hbar$ (violet solid line).
    Panels (a) and (b) are for filling $\bar{n} = 1/8$ ($L = 88$, $N = 11$), $V = 0$ and $V = J$; (c) and (d) are for filling
    $\bar{n} = 1/4$ ($L = 84$, $N = 21$), $V = 0$ and $V = J$. For all panels $\hbar\gamma = 0.05J$
    corresponding to $\hbar^2 D = 20 J^2$. The amplitude of the density oscillations are damped due to dephasing.
    Results obtained via MPS simulations with $\chi=8000$, $dt J = 0.01 \hbar$ and cut-off $\varepsilon=10^{-7}$.}\label{fig:densities} 
\end{figure}

Using this approach, we investigate the fate of the Tomonaga-Luttinger liquid as the system is subjected to dephasing, and consider 
times up to $tJ = 5 \hbar$. We first consider the evolution of the density profiles. In Fig.~\ref{fig:densities} we present
the density of the atoms $\langle \hat{n}_l\rangle$ as a function of the site $l$ both for $1/4$ filling, panels $(a)$ and $(b)$,
and filling $1/8$, panels $(c)$ and $(d)$. We consider both hard-core bosons without nearest-neighbor interaction, $V=0$ in panels $(a)$
and $(c)$, and with interaction strength $V = J$ in panels $(b)$ and $(d)$. Each line corresponds
to a different time, namely $tJ = 0$ for the blue dashed line, $tJ = 3 \hbar$ for the orange dash-dotted line
and $tJ = 5 \hbar$ for the violet solid line. Note that we only show half the system, as the other half is mirror-symmetric.
In Fig.~\ref{fig:densities}, we observe that the initial density profile has a strong oscillatory behavior. One can notice
that their amplitude is larger when the nearest-neighbor interaction is finite, and that their period is shorter for larger
fillings. The presence of these Friedel oscillations is due to the use of open-boundary conditions, and their structure at $t=0$
is in excellent agreement with the density profile expected for a finite-size
Tomonaga-Luttinger liquid. In an intermediate region slightly away from the system edge
the initial density profile follows the expected expression
\begin{eqnarray}
\langle \hat{n}_l \rangle = \frac{1}{2} + \frac{q}{2\pi} + \sqrt{2 A_z}~(-1)^l~\frac{\sin(q l)}{(2 l)^{1/(2\eta)}} \nonumber
\end{eqnarray}
where, for a finite size system, $q = 2\pi \frac{L}{L+1} \left( \bar{n} - \frac{1}{2} \right)$~\cite{HikiharaFurusaki2001},
implying that the decay of the Friedel oscillations is dominated by the Tomonaga-Luttinger exponent.

\begin{figure}[h!]
  \includegraphics[width=0.9\linewidth]{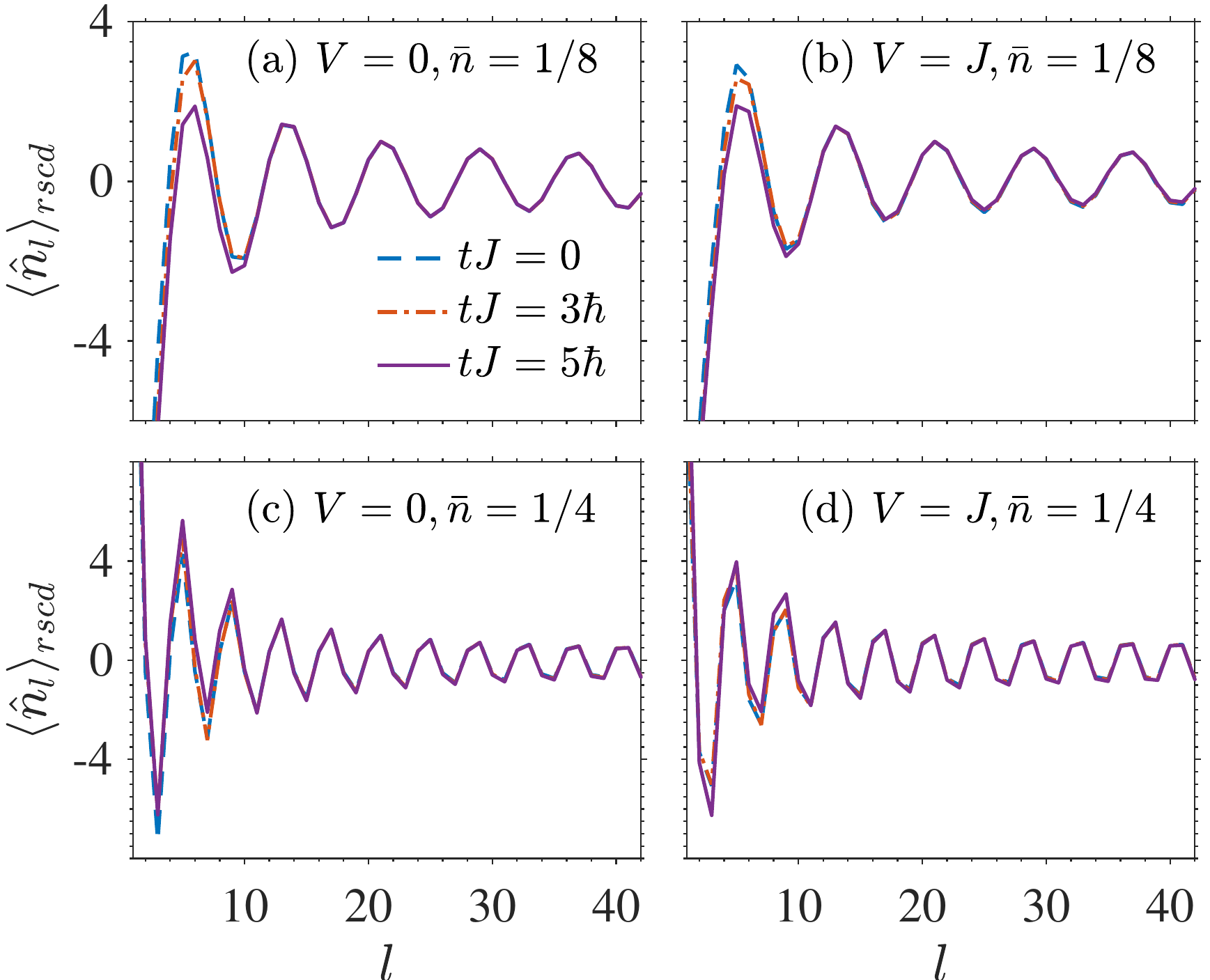}
  \caption{Rescaled local density $\langle \hat{n}_l \rangle_\text{rscd}$ versus site number $l$, for different times:
    $tJ = 0$ (blue dashed line), $tJ = 3 \hbar$ (orange dash-dotted line) and $tJ = 5 \hbar$ (violet solid line).
    Panels (a) and (b) are for filling $\bar{n} = 1/8$ ($L = 88$, $N = 11$), $V = 0$ and $V = J$; (c) and (d) are for filling
    $\bar{n} = 1/4$ ($L = 84$, $N = 21$), $V = 0$ and $V = J$. For all panels $\hbar\gamma = 0.05J$. For all system parameters
    and times considered, the bulk of the evolved profiles can be collapsed into the corresponding
    initial density profile. Results obtained via MPS simulations using the same parameters as
    in Fig.~\ref{fig:densities}.}\label{fig:densities_collapse}    
\end{figure}

As the dephasing dynamics sets in, the oscillations start to damp out, but the time-evolved density profiles appear to overall retain
their initial shape. Motivated by this observation, we analyze further whether the evolution of the density profile keeps indeed a
self-similar dependence on position. To check if this is the case, we rescale each profile within a given parameter set using the following
procedure: we subtract from each profile the density at which all three profiles of the set intersect, we then divide each newly
obtained profile by a single density, chosen from a site in the bulk of the system, for its maximal value. The result of this
simple shift and rescaling is presented in Fig.~\ref{fig:densities_collapse}. One can see that the profiles at different times
collapse onto a single curve for sites slightly away from the edges. As the initial profile is the ground state density configuration
for a Tomonaga-Luttinger liquid, our results support that the functional form predicted by Tomonaga-Luttinger
liquid theory is well preserved for times up to $J t = 5 \hbar$, although the oscillations are damped under the effect of dissipation.   

\begin{figure}[!h]
  \includegraphics[width=0.9\linewidth]{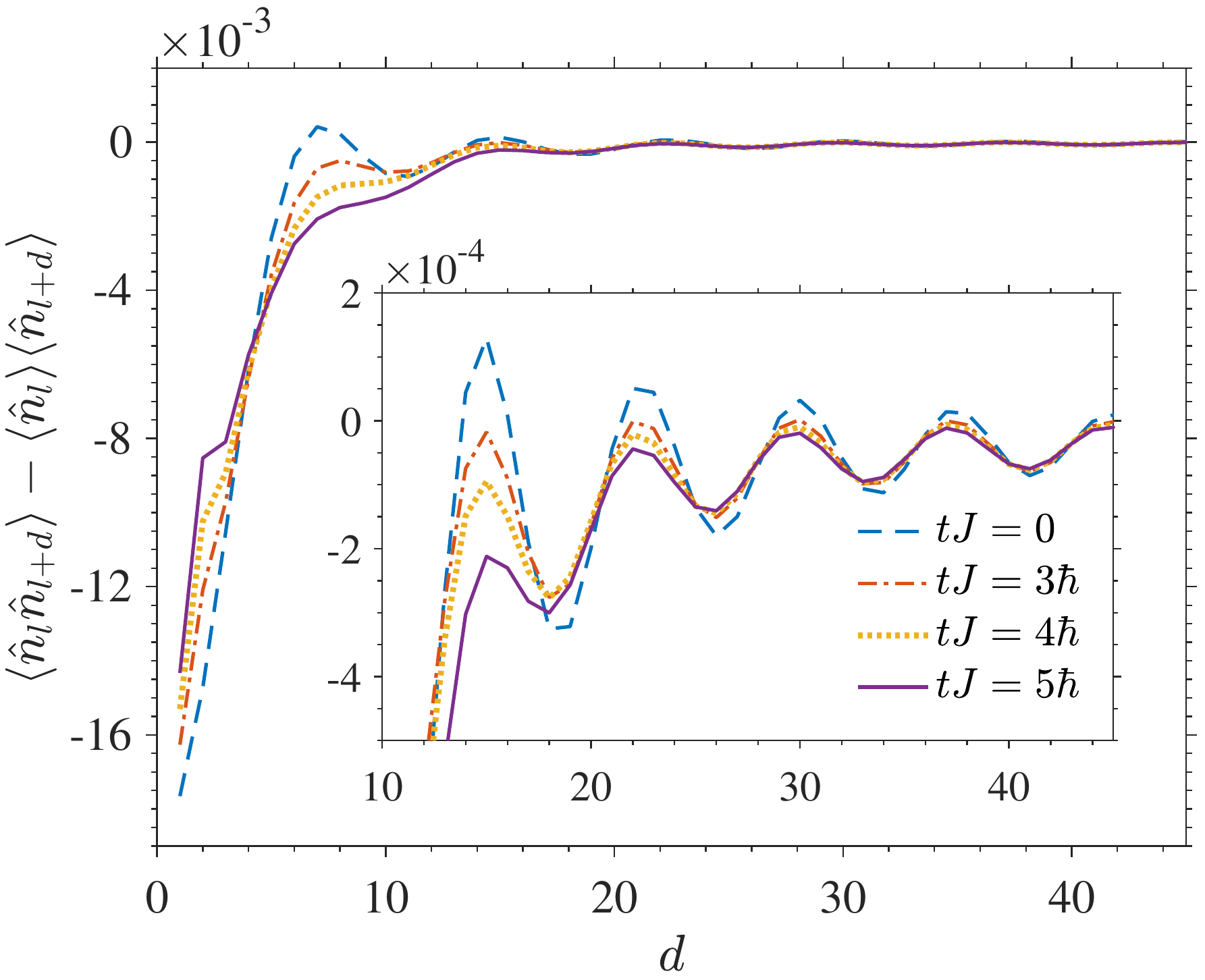}
  \caption{Density-density correlation
    $\langle \hat{n}_l \hat{n}_{l+d} \rangle - \langle \hat{n}_l \rangle \langle \hat{n}_{l+d} \rangle$
    as a function of distance $d$ ($l = 37$) for $L = 88$, $N = 11$, $V = J$ and $\hbar\gamma  = 0.05J$.
    Times considered: $tJ = 0$ (blue dashed line), $tJ = 3 \hbar$ ($D t = 60$, orange dash-dotted line),
    $tJ = 4 \hbar$ ($D t = 80$, yellow dotted line) and $tJ = 5 \hbar$ ($D t = 100$, violet solid line).
    Inset: the spatial oscillations of the Tomonaga-Luttinger correlations are damped under the effect of dephasing.
    Results obtained via MPS simulations using the same parameters as in Fig.~\ref{fig:densities}.}\label{fig:nn}
\end{figure}

To obtain a more definite picture, we also consider the evolution of the density-density correlations,
$\langle \hat{n}_l \hat{n}_{l+d} \rangle - \langle \hat{n}_l \rangle \langle \hat{n}_{l+d} \rangle$.
Just as we did in our study of the results from adiabatic elimination, we first take a look at these correlations
for short to intermediate distances. One sees here too, from Fig.~\ref{fig:nn}, that spatial oscillations
present initially in the Tomonaga-Luttinger liquid correlations are preserved for a certain time, but are
damped out under the effect of dephasing. One can further notice a slow build-up of the correlations
for short distances and hints of a breakdown of the algebraic character propagating towards larger distances
as time goes on. This behavior is clearly reminiscent of the evolution identified for the corresponding correlations
within adiabatic elimination. However, as expected, the time scales are different. While from adiabatic elimination,
the oscillations appear to have totally vanished by $D t \approx 10$, when the evolution is carried out using MPS,
properly taking in account the initial coherence and the interaction, oscillations are still present,
as seen from Fig.~\ref{fig:nn}, at $D t \approx 100$ (corresponding here to $tJ = 5\hbar$).
\begin{figure}[!h]
  \includegraphics[width=0.9\linewidth]{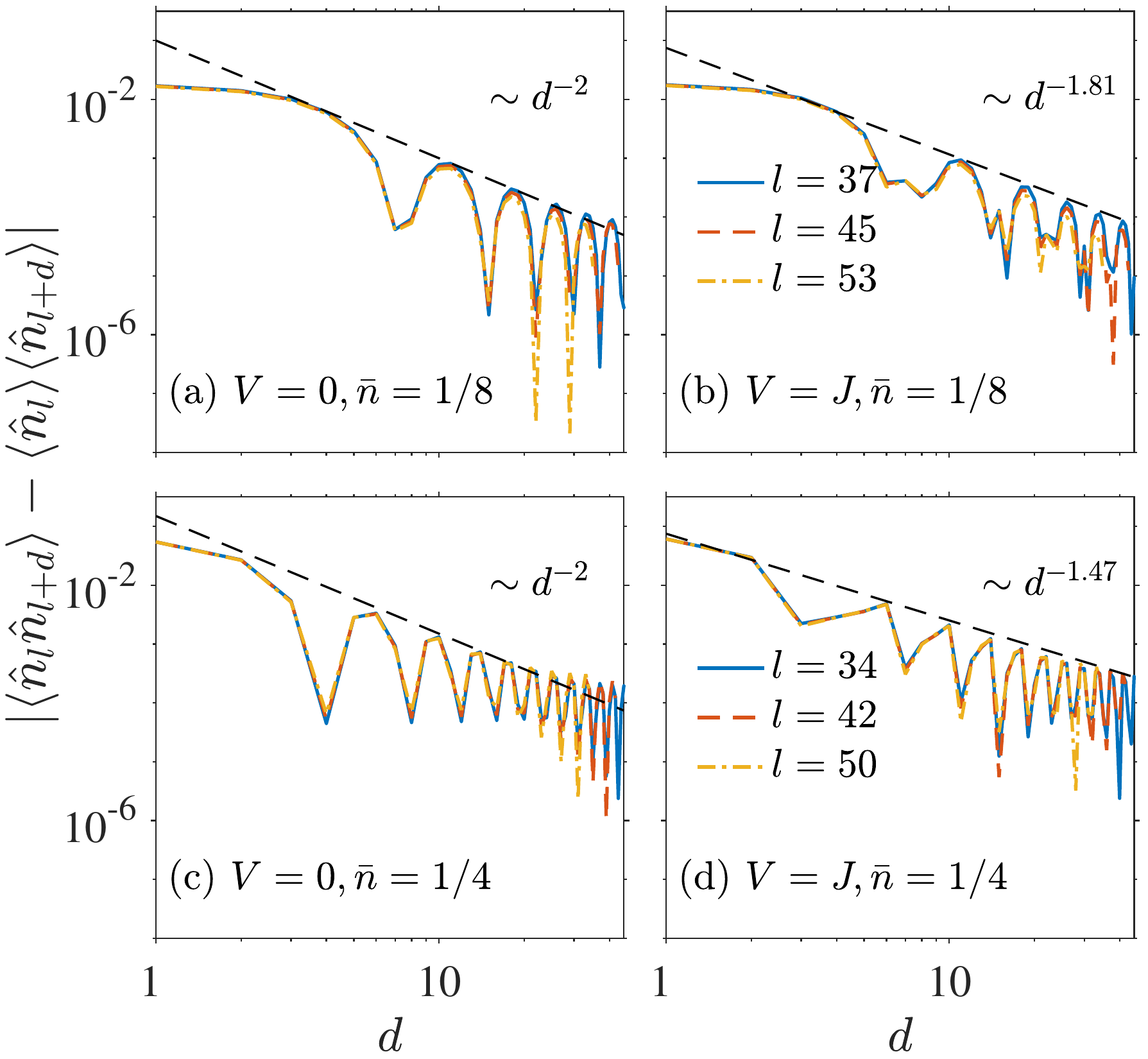}
  \caption{
    Absolute value of the initial density-density correlations
    $|\langle \hat{n}_l \hat{n}_{l+d} \rangle - \langle \hat{n}_l \rangle \langle \hat{n}_{l+d} \rangle|$
    as a function of distance $d$ for three different locations.
    Panels (a) and (b) are for filling $\bar{n} = 1/8$ ($L = 88$, $N = 11$), $V = 0$ and $V = J$; (c) and (d) are for filling
    $\bar{n} = 1/4$ ($L = 84$, $N = 21$), $V = 0$ and $V = J$.  As well
    known from Tomonaga-Luttinger theory, the exponent $\eta$ in $d^{-1/\eta}$ is function of the filling and
    interaction strength. Results obtained from MPS simulations using the same
    parameters as in Fig.~\ref{fig:densities}.}\label{fig:nn_initial_algebraic}
\end{figure}

Considerable information can be further gained by considering how the spatial algebraic decay of the
density-density correlations is affected by dephasing. Before analyzing the evolution of this decay as function
of time, we first study the structure of the initial correlations. In Fig.~\ref{fig:nn_initial_algebraic}, we
show that for various starting locations, $l$, the correlations are in very good agreement if the initial sites
correspond to a maximum of the density profile near the center of the lattice. In particular, we find that the
exponent associated with the correlations decay agrees with the value expected
from Tomonaga-Luttinger theory~\cite{HikiharaFurusaki2001}, even though for the largest distances considered
finite size effects are noticeable.

\begin{figure}[!h]
  \includegraphics[width=0.9\linewidth]{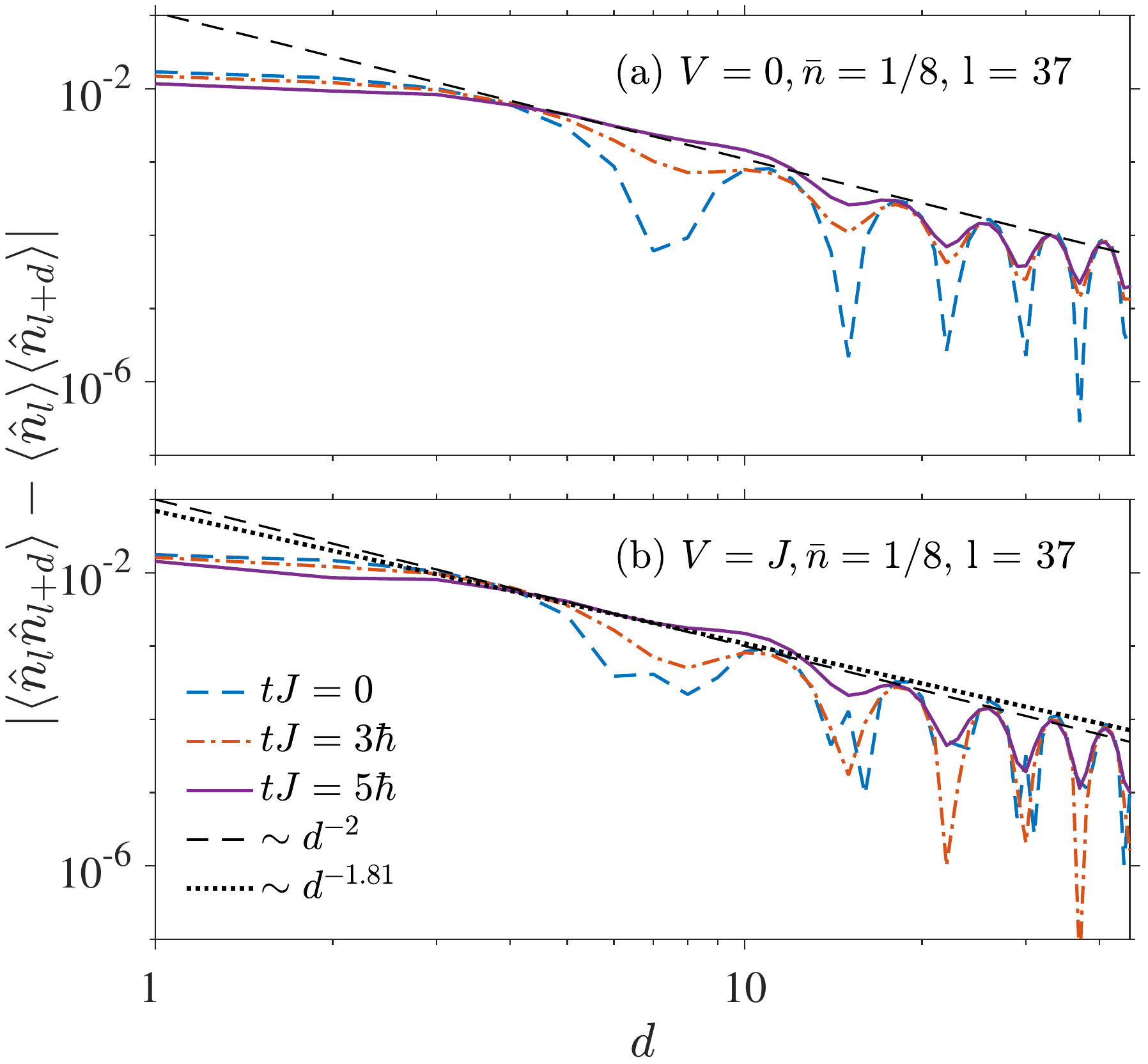}
  \caption{Absolute value of the density-density correlations
    $|\langle \hat{n}_l \hat{n}_{l+d} \rangle - \langle \hat{n}_l \rangle \langle \hat{n}_{l+d} \rangle|$
    as a function of distance $d$ for $\bar{n} = 1/8$ ($L = 88$, $N = 11$) at three times: $tJ = 0$ (blue dashed line),
    $tJ = 3 \hbar$ (orange dash-dotted line) and $tJ = 5 \hbar$ (violet solid line). Panel (a): $V = 0$, panel (b): $V = J$.
    For all panels $\hbar\gamma = 0.05J$. Results obtained via MPS simulations using the same parameters as
    in Fig.~\ref{fig:densities}.}\label{fig:nn_1ov8_algebraic_decay}
\end{figure}

\begin{figure}[!h]
  \includegraphics[width=0.9\linewidth]{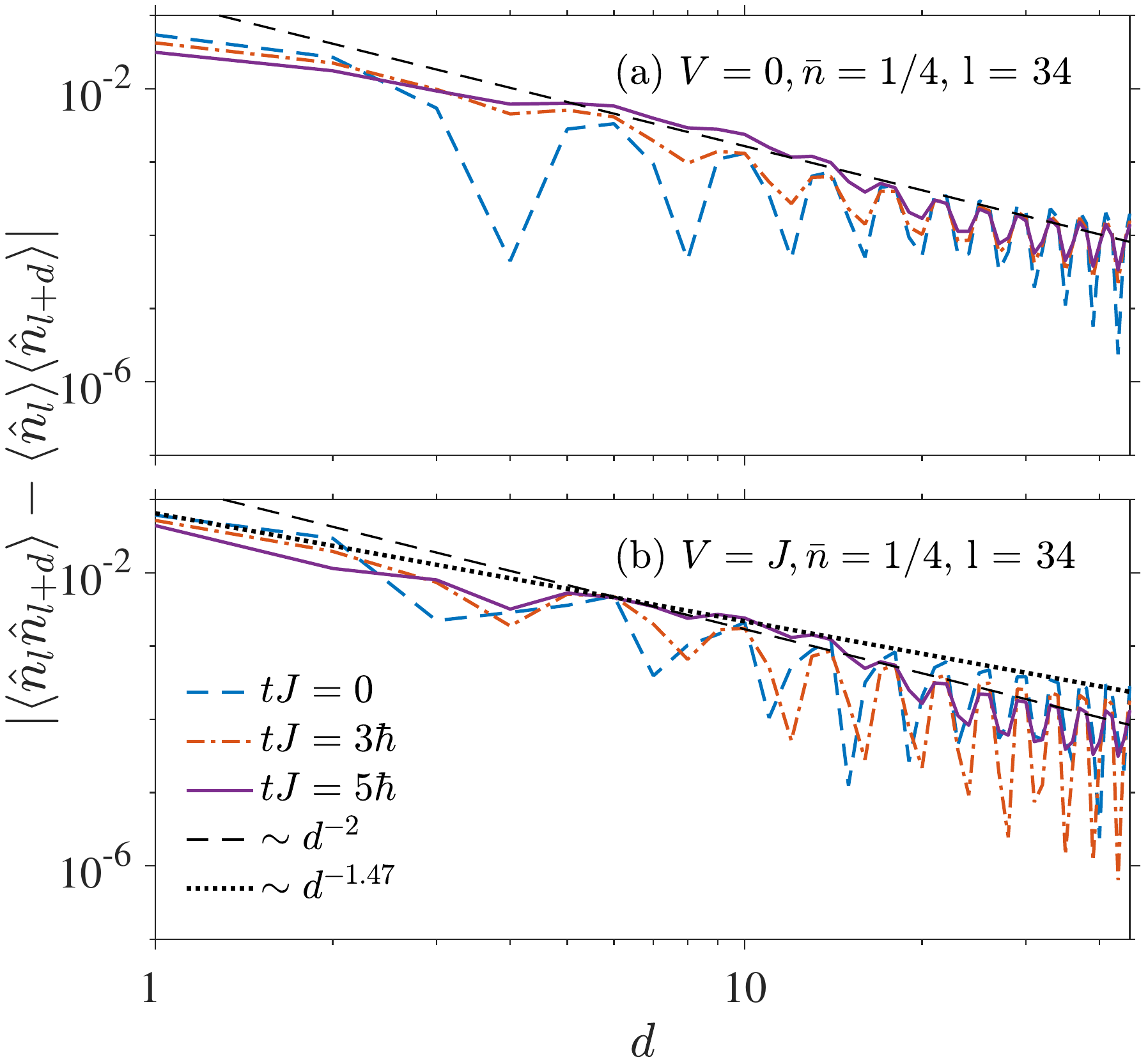}
  \caption{Absolute value of the density-density correlations
    $|\langle \hat{n}_l \hat{n}_{l+d} \rangle - \langle \hat{n}_l \rangle \langle \hat{n}_{l+d} \rangle|$
    as a function of distance $d$ for $\bar{n} = 1/4$ ($L = 84$, $N = 21$) at three times: $tJ = 0$ (blue dashed line),
    $tJ = 3 \hbar$ (orange dash-dotted line) and $tJ = 5 \hbar$ (violet solid line). Panel (a): $V = 0$, panel (b): $V = J$.
    For all panels $\hbar\gamma = 0.05J$. Results obtained via MPS simulations using the same parameters as
    in Fig.~\ref{fig:densities}.}\label{fig:nn_1ov4_algebraic_decay}
\end{figure}

We are now in a position to investigate the evolution of the tail of the density-density correlations
considering both fillings $\bar{n} = 1/8$ and $\bar{n} = 1/4$ and interaction strengths $V = 0$ and
$V = J$. In all cases displayed in Figs.~\ref{fig:nn_1ov8_algebraic_decay} and
\ref{fig:nn_1ov4_algebraic_decay}, one sees that the amplitude of the correlations flattens over time
and that this flat region grows. This result hints at the existence of the second regime
identified within adiabatic elimination, described using Eq.~(\ref{eq:cdapprox2}), where a featureless
region propagates diffusively through the system. We also notice that for larger distances the exponent
associated with the algebraic decay seems, in some cases, to change as dephasing sets in.
This is particularly noticeable in Fig.~\ref{fig:nn_1ov4_algebraic_decay}(b), for $\bar{n} = 1/4$ and $V = J$,
where the initial decay in $d^{-1.47}$ changes with time and appears to adopt the scaling form $d^{-2}$. A similar
situation likely occurs for $\bar{n} = 1/8$ and $V = J$ shown in Fig.~\ref{fig:nn_1ov8_algebraic_decay}(b); however, in this
case the initial algebraic behavior is in $d^{-1.81}$ so the evolution towards $d^{-2}$ is much harder to confidently ascertain.
In the absence of interaction, the decay remains proportional to $d^{-2}$ as predicted
by Eq.~(\ref{eq:cdapprox1}). These results are therefore in agreement with the regimes predicted from adiabatic
elimination.

\section{Conclusions} \label{sec:conclusions}  
We considered a system of interacting bosons under the effect of dephasing. The system
starts as a Tomonaga-Luttinger liquid and then, due to dephasing, heats up. We investigated the quantum
dissipative evolution of this system using an approach based on adiabatic elimination
and counterchecked our analytical findings by conducting matrix product states simulations.

We found the dephasing to act in two concurring ways: at short to intermediate times, the oscillatory
nature of both the density profile and density-density correlations damps out, but their 
overall algebraic decay, as expected for a Tomonaga-Luttinger liquid, is maintained. However,
dephasing is found to alter the interaction and filling-specific exponent associated with the initial algebraic decay.
In fact, the correlations scaling, while remaining algebraic, evolves towards the exponent expected
for a non-interacting system. Then, at larger times, the algebraic scaling diffusively melts away
and is replaced by featureless correlations as expected in the infinite temperature state. 

Our work shines a different light on the persistence of universal behaviors in strongly interacting
systems under the effect of dissipation, a difficult but important subject that will surely benefit greatly
from further refinements of both the theoretical tools and experimental setups in coming years. 

\section*{Acknowledgement}
We thank V. Balachandran, M. Buchhold, S. Diehl, G. Kocher and J. Marino for fruitful and enlightening discussions.      
The computational work for this article was partially done on resources of the National
Supercomputing Centre, Singapore \cite{nscc}. D.P. acknowledges support from Ministry of Education of
Singapore AcRF MOE Tier-II (project MOE2018-T2-2-142).
C.K. acknowledges funding from the Deutsche Forschungsgemeinschaft (DFG, German Research Foundation)
under project number 277625399 - TRR 185 (B4) and project number 277146847 - CRC 1238 (C05) and under
Germany's Excellence Strategy – Cluster of Excellence Matter and Light for
Quantum Computing (ML4Q) EXC 2004/1 – 390534769 and the European Research Council (ERC)
under the Horizon 2020 research and innovation programme, grant agreement No.~648166 (Phonton).


\bibliography{bernier_luttinger_dephasing}
  
\end{document}